\documentclass[preprint,prl,showpacs,floatfix,superscriptaddress]{revtex4-1}
\usepackage{amsmath, amssymb}
\usepackage{times}
\usepackage {euscript}
\usepackage{graphicx}
\usepackage{amsthm}
\usepackage{epsfig}
\usepackage{bm}
\usepackage{multirow}
\usepackage{longtable}
\usepackage[dvipsnames]{xcolor}
\newcommand{\bp}{{\bf p}}
\newcommand{\bk}{{\bf k}}
\newcommand{\br}{{\bf r}}

\newcommand{\balpha}{\boldsymbol{\alpha}}
\newcommand{\bepsilon}{\boldsymbol{\epsilon}}

\graphicspath{{graph/}}
\begin{document}
\thispagestyle{empty}
\title{
Ab initio QED treatment of the two-photon annihilation of positrons with bound electrons
}
\author{V.~A.~Zaytsev}
\affiliation{
Department of Physics, St. Petersburg State University,
Universitetskaya naberezhnaya 7/9, 199034 St. Petersburg, Russia
}
\author{A.~V.~Volotka}
\affiliation{
Department of Physics, St. Petersburg State University,
Universitetskaya naberezhnaya 7/9, 199034 St. Petersburg, Russia
}
\affiliation{
Helmholtz-Institut Jena, Fr\"obelstieg 3, D-07743 Jena, Germany
}
\affiliation{
GSI Helmholtzzentrum f\"ur Schwerionenforschung, D-64291 Darmstadt, Germany
}
\author{D.~Yu}
\affiliation{
Institute of Modern Physics, Chinese Academy of Sciences, 
Lanzhou 730000, China
}
\author{S.~Fritzsche}
\affiliation{
Helmholtz-Institut Jena, Fr\"obelstieg 3, D-07743 Jena, Germany
}
\affiliation{
GSI Helmholtzzentrum f\"ur Schwerionenforschung, D-64291 Darmstadt, Germany
}
\affiliation{
Theoretisch-Physikalisches Institut, 
Friedrich-Schiller-Universit\"at Jena, Jena D-07743, Germany
}
\author{X.~Ma}
\affiliation{
Institute of Modern Physics, Chinese Academy of Sciences, 
Lanzhou 730000, China
}
\author{H.~Hu}
\affiliation{
Hypervelocity Aerodynamics Institute, China Aerodynamics Research and Development Center, 
621000 Mianyang, Sichuan, China
}
\author{V.~M.~Shabaev}
\affiliation{
Department of Physics, St. Petersburg State University,
Universitetskaya naberezhnaya 7/9, 199034 St. Petersburg, Russia
}
%
\begin{abstract}
The process of a positron -- bound-electron annihilation with simultaneous 
emission of two photons is investigated theoretically.
A fully relativistic formalism based on {\it ab initio} QED description of 
the process is worked out.
The developed approach is applied to evaluate the annihilation of a positron with $K$-shell electrons of a silver atom, for which a strong contradiction between theory and experiment was previously stated.
The results obtained here resolve this long-standing disagreement and, moreover, demonstrate
a sizeable difference with approaches so far used for calculations of the positron -- bound-electron annihilation process, namely, the {\it Lee's} and {\it impulse} approximations.
\end{abstract}
%
\maketitle
%
%
Since the first observation of positrons~\cite{Anderson_PR41_405:1932}, 
investigations of their interaction with atoms, molecules, and solids are of 
unaltered interest (see, e.g., Refs.~\cite{Iwata_PRL79_39:1997, 
Weber_PRL83_4658:1999, Hunt_PRL86_5612:2001, Cizek_NJP14_035005:2012} and the 
review~\cite{Surko_JPB38_R57:2005}).
Extensive investigations of the positron annihilation processes gave rise to numerous 
applications ranging from astrophysical researches~%
\cite{Guessoum_AA436_171:2005, Lingenfelter_PRL103_031301:2009, 
Prantzos_RMP83_1001:2011} and positron-induced Auger-electron 
spectroscopy~\cite{Weiss_RPC76_285:2007, Mayer_PRL105_207401:2010}
to studies of the defects in metals and semiconductors~%
\cite{Weiss, Tuomisto_RMP85_1583:2013}, dynamics of catalysis~%
\cite{Mayer_PRL105_207401:2010} and positron-emission tomography~%
\cite{PET:2005,PET:2006}.
In particular, the angular distribution of the photon pairs from the annihilation defines the spatial resolution of the defect analysis and tomography.
Apart from the various applications, investigations of the positron annihilation
with inner-shell electrons of heavy ionic (or atomic) targets can give a valuable
insight into the kinematically analogous meson decays in quantum chromodynamics (QCD).
In these decays, the QCD coupling constant is mimicked by the effective 
electromagnetic coupling constant being enlarged by the nuclear charge~%
\cite{Greub_PRD52_4028:1995, Brodsky_HI209_83:2012}.
Therefore, a quantitative understanding of the positron -- bound-electron annihilation is
highly requested by the on-going growth of studies considering positron -- matter interaction as
well as by upcoming positron facilities of a new generation, e.g., at Lawrence Livermore National
Laboratory \cite{Chen_PRL102_105001:2009, Chen_PRL105_115003:2010, Piazza_RMP84_1177:2012,
Sarri_PRL110_255002:2013} and ELI-NP Research Centre \cite{Hugenschmidt_APB106_241:2012,
Djourelov_NIMA806_146:2016}.
%
%
\\ 
\indent
%
%
The positron -- bound-electron annihilation can proceed with the emission of 
one, two, or even more photons. 
More often than not, the two-quantum annihilation dominates over other channels. 
This process, however, has not been described rigorously within the framework of QED and with a proper account of the interaction with a nucleus yet.
So far the calculations of the positron -- bound-electron two-quantum 
annihilation were just based on two approximations:
{\it Lee's} approach~\cite{Chang_ZETF33_365:1957} for ultra-slow (thermalized) and the {\it impulse} approximation for ultra-fast positrons.
For slow positrons, the dominant contribution to the overall 
annihilation cross section with atomic targets arises from the {\it nonrelativistic} 
valence and outer shell electrons. 
These processes can be well described in the framework of the nonrelativistic Lee's~\cite{Chang_ZETF33_365:1957} approximation.
On the basis of this approximation the theoretical approach which shows a 
remarkable agreement with related experimental studies was developed~
\cite{Gribakin_PRA70_032720:2004, Dunlop_JPB39_1647:2006, 
Green_PRA90_032712:2014, Green_PRL114_093201:2015}.

\begin{figure}[h!]
\includegraphics[width=0.48\textwidth]{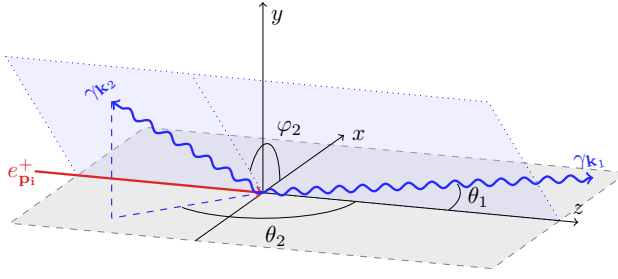}
\caption{
Geometry (in the ion rest frame) of the positron ($e^+_{\bp_i}$) -- bound-electron
annihilation with the emission of two photons $\gamma_{\bk_1}$ and $\gamma_{\bk_2}$.
}
\label{fig:geometry}
\end{figure}

\begin{figure}[h]
\includegraphics[width=0.48\textwidth]{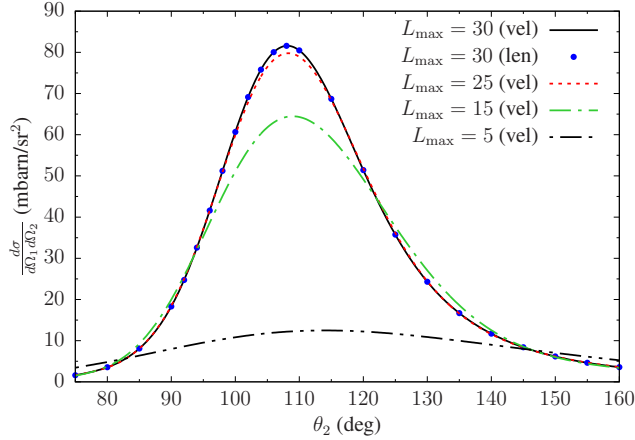}
\caption{
Double differential angular cross section~\eqref{eq:d2cross} for the 
two-quantum annihilation of the $300$ keV positron with the $K$-shell electrons 
of a silver atom for different numbers of the photon multipoles $L_{\rm max}$ 
taken into account.
The emission angles $\theta_1 = 30^\circ$ and $\varphi_2 = 180^\circ$.
}
\label{fig:d2cross_mult}
\end{figure}

\begin{figure}[h]
\includegraphics[width=0.48\textwidth]{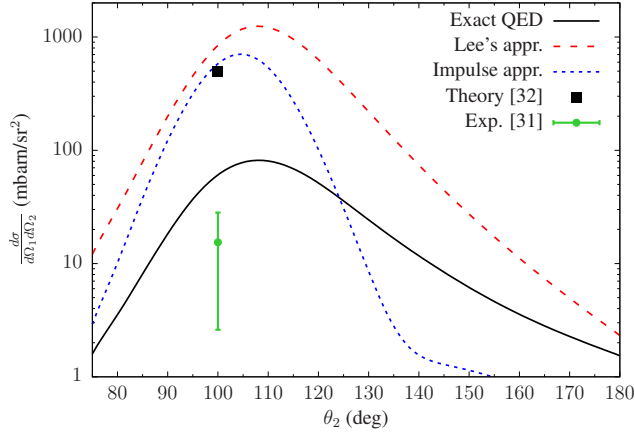}
\caption{
Double differential angular cross section~\eqref{eq:d2cross} for the 
two-quantum annihilation of the 300 keV positron with the $K$-shell electrons 
of a silver atom.
The calculations performed within the exact approach, Lee's, and 
impulse approximations are represented by the black solid, red dashed, and blue 
dotted lines, respectively.
The theoretical result from Ref.~\cite{Gorshkov_JETP45_17:1977} is shown by a 
black square and experimental value~\cite{Nagatomo_PRL32_1158:1974} is depicted 
by a green circle with error bars.
The logarithmic scale is chosen for the $y$-axis.
}
\label{fig:d2cross}
\end{figure}

On the basis of this approximation, the theoretical approach which shows a remarkable agreement with related experimental studies was developed~\cite{Gribakin_PRA70_032720:2004, Dunlop_JPB39_1647:2006, 
Green_PRA90_032712:2014, Green_PRL114_093201:2015}.
For \textit{ultra-fast} positrons, in contrast, the impulse approximation can be applied, in which all particles are assumed to be free and where the active electron is represented by a stationary wave packet of superimposed plane-wave states.
In this approximation, the annihilation process is based on the formulas which were derived almost a century ago by Dirac~\cite{Dirac_ZP62_545:1930} and Tamm~\cite{Tamm_ZP62_545:1930}.
However, these two approximations cannot be applied to the annihilation of positrons with inner-shell electrons and for collision energies, at which the interaction with a nucleus plays a significant role.
As an example, we refer to the experiment, where the two-quantum annihilation 
of $300$ keV positrons with $K$-shell electrons of silver was measured~%
\cite{Nagatomo_PRL32_1158:1974}, and for which the theoretical cross sections by Gorshkov and coworkers~\cite{Gorshkov_JETP45_17:1977, Drukarev} differ by more than an order of magnitude.
%
%
\\ 
\indent
%
%
Here, we develop a fully-relativistic formalism based on {\it ab initio} QED
description of the two-quantum annihilation of positrons with bound electrons.
In this formalism, positron- and electron-nucleus interaction is treated nonperturbatively.
As the first application, we use the developed approach for the description
of the two-quantum annihilation of positrons with $K$-shell electrons of silver which was studied
experimentally in Ref.~\cite{Nagatomo_PRL32_1158:1974}.
Our double differential angular cross section of $41(12)$ mbarn/sr$^2$ is in good 
agreement with the experimental value $15.4(12.8)$ mbarn/sr$^2$, and which resolve long-standing disagreement between theory and experiment.
Additionally, we compare the results of the developed exact approach with ones 
obtained within the Lee's and impulse approximations as well as with result of 
Ref.~\cite{Gorshkov_JETP45_17:1977} and discuss possible reasons of the 
discrepancies.
%
%
\\ \indent
%
%
The differential cross section for the two-quantum annihilation of a positron 
with a bound electron in the relativistic units $\hbar=1$, $c=1$, $m=1$ is given by~\cite{Akhiezer, Berestetsky}
\begin{equation}
\frac{d\sigma}{d\bk_1 d\bk_2} = 4\alpha^2 
\frac{(2\pi)^6}{v_i} \left\vert \tau \right\vert^2 
\delta(E_a + \varepsilon_i - \omega_1 - \omega_2),
\label{eq:cross}
\end{equation}
where $\alpha$ is the fine structure constant, $\varepsilon_i$ and $v_i$ are the 
energy and velocity of the positron, respectively, $E_a$ is the energy of 
the active electron, and $\tau$ is the amplitude whose explicit form will be 
specified below.
In the present letter, we will consider only the double differential angular cross section defined by
\begin{equation}
\frac{d\sigma}{d\Omega_1 d\Omega_2} = 
\int d\omega_1 d\omega_2  \omega_1^2 \omega_2^2 
\frac{d\sigma}{d\bk_1 d\bk_2}.
\label{eq:d2cross}
\end{equation}
This cross section is assumed to be averaged over the angular momentum and spin 
projections of the electron and positron, respectively, and summed over the 
polarizations of the emitted photons.
The solid angles of the emitted photons $\Omega_{1,2}$ are defined by the 
azimuthal $\varphi_{1,2}$ and polar $\theta_{1,2}$ angles (see Fig.~\ref{fig:geometry}).
Here the $x-z$ plane is spanned by the momenta of the incoming positron $\bp_i$
and one of the emitted photons $\bk_1$ with the $z$-axis fixed along the 
direction of $\bp_i$.
Here, we utilize the independent-particle approximation, in which the positron and
the active electron move in an effective (Coulomb and screening) potential created by the
nucleus and all the other electrons.
The screening potential is induced by the Hartree charge density of these remaining
electrons.
Based on our prior analysis for the Rayleigh scattering of high-energetic photons~\cite{volotka2016}, we expect that the independent particle approximation stays valid for the processes
involving inner-shell electrons of heavy systems, where the correlation effects are
suppressed by a factor $1/Z$ ($Z$ is the nuclear charge number).
%
%
\\ \indent
%
%
The amplitude of the two-quantum annihilation of the positron with the electron 
in the bound $a$ state is given by two Feynman diagrams shown in 
Fig.~\ref{fig:feynman}, which correspond to the following expression~%
\cite{Akhiezer, Berestetsky}:
\begin{figure}[h]
\includegraphics[width=0.48\textwidth]{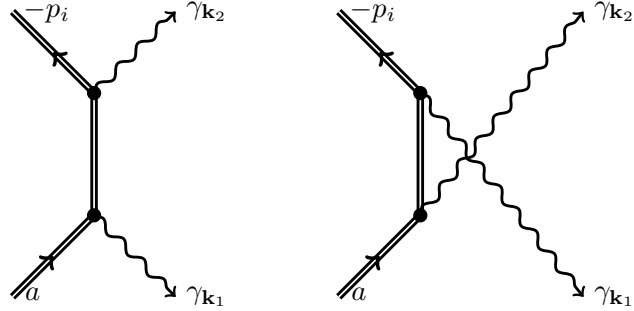}
\caption{
Feynman diagrams for the two-photon annihilation of the positron $e^+_{\bp_i}$ 
with the bound electron in the $a$ state. The double lines indicate the 
electron-positron propagators and wave functions in the external field of the 
nucleus and remaining electrons, while the wavy lines represent the emitted photons, 
$\gamma_{\bk_1}$ and $\gamma_{\bk_2}$.
}
\label{fig:feynman}
\end{figure}
\begin{widetext}
\begin{equation}
\tau = -
\sum_n\left[
\frac{\left\langle (-p_i\mu_i)
\left\vert \balpha \cdot {\bf A}^*_{\bk_2\lambda_2} \right\vert 
n\right\rangle
\left\langle n 
\left\vert \balpha \cdot {\bf A}^*_{\bk_1\lambda_1} \right\vert 
a \right\rangle}{E_a - \omega_1 - E_n(1 - i0)}
+
\frac{\left\langle (-p_i\mu_i)
\left\vert \balpha \cdot {\bf A}^*_{\bk_1\lambda_1} \right\vert 
n\right\rangle
\left\langle n 
\left\vert \balpha \cdot {\bf A}^*_{\bk_2\lambda_2} \right\vert 
a\right\rangle}{E_a - \omega_2 - E_n(1 - i0)}
\right].
\label{eq:amplitude}
\end{equation}
\end{widetext}
Here $\sum_n$ implies the complete summation over the whole spectrum, including the
integration over the positive and negative continuum parts, $\mu_i$ is the helicity of the 
incoming positron, $\balpha$ is the vector of Dirac matrices, and the wave function 
of the plane-wave photon with the polarization $\lambda$ is given by
\begin{equation}
{\bf A}_{\bk\lambda} \equiv {\bf A}_{\bk\lambda}(\br) = 
\frac{\bepsilon_\lambda e^{i\bk\cdot\br}}{\sqrt{2\omega(2\pi)^3}}.
\end{equation}
The amplitude~\eqref{eq:amplitude} determines the differential cross
section~\eqref{eq:cross} uniquely and, thus, describes the two-quantum annihilation process completely.
Let us turn to the details of the calculation of this amplitude.
%
%
\\ \indent
%
%
The infinite summation $\sum_n$ in Eq.~\eqref{eq:amplitude} is replaced
by a sum over a quasi-complete set of the Dirac equation solutions.
These solutions are obtained by using the dual-kinetic-balance finite 
basis set method~\cite{Shabaev_PRL93_130405:2004} with basis functions 
constructed from $B$ splines~\cite{Johnson_PRA37_307:1988, Sapirstein_JPB29_5213:1996}.
Such an approach yields the wave functions of the quasi-states $n$, including the bound 
state $a$, but it can barely be applied for constructing the wave function of a positron 
with a given energy.
The incoming positron with the four-momentum $p_i$ and the helicity $\mu_i$ is 
treated as an outgoing electron with the four-momentum $-p_i$ and the helicity 
$\mu_i$~\cite{Bjorken, Itzykson}.
The explicit form of the wave function of such a particle can be found, e.g., in 
Refs.~\cite{Rose, Artemyev_PRA79_032713:2009}.
The numerical construction of this wave function is performed with the use of the
modified RADIAL package~\cite{Salvat1995}.
We note that the constructed wave functions of the incoming positron, 
quasi-states $n$, and initial bound state $a$ take into account the interaction with the 
effective (Coulomb and screening) potential to all orders.
To calculate the matrix elements, we utilize the well-known multipole 
expansion technique.
As a result, one gets the infinite multipole summations over the photon and positron
multipoles, which are further restricted by analyzing the convergence property.
More details of the developed method will be presented in a forthcoming publication.
%
%
\\ \indent
%
%
As the first application of our {\it ab initio} approach, we calculate the 
double differential angular cross section (DDACS) for the process of the two-quantum
annihilation of the $300$ keV positron with the $K$-shell electrons of a silver atom
($Z = 47$), which was experimentally investigated in Ref.~\cite{Nagatomo_PRL32_1158:1974}.
In this experiment, the annihilation photons were detected at solid angles given 
by $\theta_2 = 100^\circ$, $\varphi_2 = 180^\circ$, and $\theta_1 = 30^\circ$. 
Fig.~\ref{fig:d2cross_mult} presents the DDACS as a function of the angle $\theta_2$ 
for $\theta_1$ and $\varphi_2$ being fixed as in Ref.~\cite{Nagatomo_PRL32_1158:1974}.
This figure also shows the convergence of the DDACS with respect to the number
of the photon multipoles $L_{\rm max}$ that need to be taken into account 
in the expansion of the photons wave function.
About $30$ multipoles are sufficient to obtain well-converged differential cross sections, giving rise to $60$ partial waves in the decomposition of the positron wave function.
We performed all computations both, in length and velocity gauges, and obtained perfect agreement as seen from Fig.~\ref{fig:d2cross_mult}. 
Moreover, the differential cross sections differ by less than 1\% if other than the Hartree screening potential is applied.
%
%
\\ \indent
%
%
We can also compare our {\it ab initio} QED results with those from the Lee's and impulse approximations.
In Lee's approximation~\cite{Chang_ZETF33_365:1957}, which has been widely used for the description of the two-quantum annihilation of slow (thermalized) positrons~\cite{Gribakin_PRA70_032720:2004, Dunlop_JPB39_1647:2006, Green_PRA90_032712:2014, Green_PRL114_093201:2015}, (i) the Dirac-Coulomb propagator is replaced by a free electron one,
(ii) the binding energy of the initial electron $a$ and the kinetic energy of the
incoming positron are assumed to be much smaller than the electron rest mass, and
(iii) the Dirac electron and positron wave functions are replaced by the
corresponding two-component Schr\"odinger-Pauli wave functions.
Making use of these assumptions in Eq.~\eqref{eq:amplitude}, one can obtain the 
expression for the two-quantum annihilation amplitude~\cite{Chang_ZETF33_365:1957}
\begin{equation}
\tau^{\rm (Lee)}  =  \frac{i}{2}
\left(\bk_1 - \bk_2\right)\cdot
\left\langle (-\bp_i\mu_i)^{\rm (SP)} \left\vert 
\left[{\bf A}^*_{\bk_2\lambda_2} \times {\bf A}^*_{\bk_1\lambda_1}\right]
\right\vert a^{\rm (SP)} \right\rangle.
\end{equation}
Here the superscript (SP) stands for Schr\"odinger-Pauli wave functions.
Fig.~\ref{fig:d2cross} compares the DDACS from this approximation with our {\it ab initio} results and shows that Lee's approximation overestimates the DDACS by an order of magnitude when compared with the rigorous QED prediction.
This discrepancy mainly arises from the large ($300$ keV) kinetic energy of the positron and the importance of the binding and the relativistic effects for the inner-shell electrons of a silver atom.
%
%
\\ \indent
%
%
In the relativistic impulse approximation (IA), it is assumed that the interaction with a nucleus can be neglected for high-energetic positrons and that the process can be viewed as a free positron annihilation with a stationary wave packet of superimposed plane-wave electron states.
Following the derivation which has been previously described in detail for the Compton scattering~\cite{Ribberfors_PRB12_2067:1975, Bergstrom_PRA48_1134:1993}, one can obtain the DDACS for the two-quantum annihilation in the IA
\begin{equation}
\frac{d\sigma^{\rm (IA)}}{d\Omega_1 d\Omega_2} = 
\frac{1}{v_i}
\int d\bp \rho_a(\bp)
\frac{dW_{\rm free}}{d\Omega_1 d\Omega_2}(\bp).
\label{eq:cross_ia}
\end{equation}
Here $\rho_a(\bp)$ is the momentum distribution of the initial bound-electron $a$ state and $\frac{dW_{\rm free}}{d\Omega_1 d\Omega_2}$ is the double differential angular probability for the two-quantum annihilation of a free positron and a free electron with the momentum $\bp$~\cite{Akhiezer}.
The DDACS being calculated in the IA is compared with other calculations in Fig.~\ref{fig:d2cross}.
From this figure, it is seen that the IA, like Lee's approximation, \textit{overestimates} the DDACS by an order of magnitude. 
This can be understood by the neglected interaction between the positron and the nucleus and, hence, an (unphysically) increased overlap of the positron and electron densities.
That, in turn, leads to the growth of the cross section. 
%
%
\\ \indent
%
%
Finally, we compare the obtained results with the previous theoretical 
predictions by Gorshkov and coworkers~\cite{Gorshkov_JETP45_17:1977}, which is displayed in Fig.~\ref{fig:d2cross} by a black square. 
These authors started from the free-particle approximation for the two-quantum annihilation and have evaluated the corrections of the first order in the interaction with the nucleus.
This approach corresponds to the expansion in powers of $\alpha Z$ and $\alpha Z / v$ which in the case under investigation approximately equal $0.34$ and $0.44$, respectively.
The significant deviation from the exact treatment, however, indicates that such a perturbation expansion fails to describe the DDACS of the considered process.
%
%
\\ \indent
%
%
Fig.~\ref{fig:d2cross} compares the different theoretical predictions for the DDACS with the experimental value~\cite{Nagatomo_PRL32_1158:1974}. 
First, let us note the extra factor $2$ in the denominator of Eq.~(2) in
Ref.~\cite{Nagatomo_PRL32_1158:1974}. 
This factor should appear if the contributions of the same quantum states are 
accounted twice, which does not apply for the the DDACS. 
Therefore, here and below the results from Ref.~\cite{Nagatomo_PRL32_1158:1974} 
are multiplied by a factor $2$. 
From Fig.~\ref{fig:d2cross} it is seen that all approximate theoretical results, 
including that of Ref.~\cite{Gorshkov_JETP45_17:1977}, are by an order of 
magnitude away from the experimental value, and quite in contrast to our 
rigorous QED treatment that provides the prediction which is rather close to 
the experimental result.
%
%
\\ \indent
%
%
However, the \textit{direct} comparison of the calculated DDACS with the experimental value might not be fully justified in Fig.~\ref{fig:d2cross}. 
This is caused by the fact that the experimental value just represents an detector-averarged DDACS. 
In Ref. [31], indeed, the weighted averages of the overall detector efficiencies, including the geometrical factors, $(\epsilon\Omega)_1$ and $(\epsilon\Omega)_2$ for the annihilation photons
are defined to be $6.5\times10^{-2}$~sr and $5.8\times10^{-2}$~sr, respectively.
If we assume a 100\% efficiency for both detectors, we can evaluate the 
detector-averaged DDACS within the exact QED approach as well as within the Lee's 
and impulse approximations. 
In the case of the exact calculation, $15$ photon multipoles were taken into 
account. 
The integration over each photon emission angle is performed by a 4-point 
Gauss-Legendre quadrature. 
A conservative error estimate for the averaged DDACS gives 30\%. 
In Table~\ref{tb:comparison} we compare the detector-averaged DDACS calculated 
within the exact approach, the Lee's and impulse approximations with the 
experimental value~\cite{Nagatomo_PRL32_1158:1974}.
\begin{table}[h]
\caption{
Detector-averaged double differential angular cross section for the two-quantum 
annihilation of the $300$ keV positron with the $K$-shell electrons of a silver atom.
The experimental value from Ref.~\cite{Nagatomo_PRL32_1158:1974} is multiplied by a
factor $2$ (see the text for details).
}
\tabcolsep10pt
\begin{tabular}{cc} \hline\hline
Approach & Detector-averaged DDACS (mbarn/sr$^2$) \\
\hline
Impulse appr.                               & $240$ \\
Lee's appr.                                 & $300$ \\
Exact QED                                   & \;\;\;\;\;\;\;\;\;$41(12)$ \\
\hline
Experiment~\cite{Nagatomo_PRL32_1158:1974}       & \;\;\;\;\;\;\;\;\;\;\;\;\;\;\;$15.4(12.8)$\\ \hline\hline
\end{tabular}
\label{tb:comparison}
\end{table}
From the table, one can see that after the averaging of the DDACS decreases by almost a
factor 2 because the planar geometry and detector position at $\approx 180^\circ$ just refers to the maximum of the cross section. 
Any deviation from this geometry leads to the drop of the DDACS.
Here it is worth noting that smaller detector efficiencies will lead to the further decrease of the DDACS.
From table~\ref{tb:comparison}, one can also see that the results of {\it ab initio} QED approach are in good agreement with the experimental value.
%
%
\\ \indent
%
%
In conclusion, a fully relativistic QED description of the 
two-quantum annihilation of a positron with a bound electron is presented 
for the very first time.
This novel approach has been applied for the annihilation of $300$ keV positrons with the $K$-shell electrons of silver. 
Our result for the double differential angular cross section is in good 
agreement with the experimental value~\cite{Nagatomo_PRL32_1158:1974} and, thus, resolves long-standing disagreement between theory and experiment.
It was also shown that none of the approaches so far used for the
calculation of the positron -- bound-electron two-quantum annihilation can be 
applied in this case.
We believe that the exact approach developed here can be extended to many other cases of the positron annihilation and, thus, will allow to establish more precise validity criteria for the so far employed approximations as well as to help in the interpretation of the experimental data in various applications of the positron annihilation processes.

%
%
\begin{acknowledgments}
This work was supported by RFBR-NSFC (Grants No. 17-52-53136 and No. 11611530684), by SPbSU-DFG (Grants No. 11.65.41.2017 and No. STO 346/5-1), and by DFG (Grant No. VO1707/1-3).
A.V.V. and V.M.S. acknowledge the support of the CAS President International Fellowship Initiative (PIFI).
H.H. acknowledges the support from the NSFC (Grant No. 11774415).
The work of V.A.Z. was also supported by the grant of the President of the Russian Federation (Grant No. MK-4468.2018.2) and by SPbSU (TRAIN 2018: 34825432).
\end{acknowledgments}
%
%

%
\end {document}